# Optical Imaging of Surface Chemistry and Dynamics in Confinement

Carlos Macias-Romero, Igor Nahalka, Halil I. Okur, Sylvie Roke[#]

*Laboratory for fundamental BioPhotonics (LBP), Institute of Bioengineering (IBI), and Institute of Materials Science (IMX), School of Engineering (STI), and Lausanne Centre for Ultrafast Science (LACUS), École Polytechnique Fédérale de Lausanne (EPFL), CH-1015, Lausanne, Switzerland,* [#]*sylvie.roke@epfl.ch;*

**Surface chemistry and dynamics are optically imaged label-free and non-invasively in a 3D confined geometry and on sub-second time scales**

**Abstract**

The interfacial structure and dynamics of water in a microscopically confined geometry is imaged in three dimensions and on millisecond time scales. We developed a 3D wide-field second harmonic microscope that employs structured illumination. We image pH induced chemical changes on the curved and confined inner and outer surfaces of a cylindrical glass micro-capillary immersed in aqueous solution. The image contrast reports on the orientational order of interfacial water, induced by charge-dipole interactions between water molecules and surface charges. The images constitute surface potential maps. Spatially resolved surface $pK_{a,s}$ values are determined for the silica deprotonation reaction. Values range from $2.3 < pK_{a,s} < 10.7$, highlighting the importance of surface heterogeneities. Water molecules that rotate along an oscillating external electric field are also imaged. With this approach, real time movies of surface processes that involve flow, heterogeneities and potentials can be made, which will further developments in electrochemistry, geology, catalysis, biology, and microtechnology.

Microscopic and nanoscopic structural heterogeneities, confinement, and flow critically influence surface chemical processes in electrochemical, geological, and catalytic reactions (*1-5*). Recent advances in micro- and nanotechnology are driven by these phenomena, often in combination with intrinsic surface or externally applied electrostatic field gradients. Micro- and nanocapillaries in combination with interfacial voltage gradients are used in the fabrication, manipulation, and characterization of droplets and interfaces (*6-9*) and highly amorphous materials (*10*). An electrostatic field gradient in a narrow channel can be used to separate (*11*) and identify analytes (*12*). Naturally occurring pores and interfaces employ extremely high electric field gradients to separate and transport chemicals from one aqueous phase to another (*13*). Artificial nanopores have recently been used to generate liquid jets (*14*) and electric power (*15, 16*). The increase in complexity of new materials and nano- and microtechnological developments (*17*) requires technology that can track, in real time, three dimensional spatial changes in the molecular structure of confined systems, such as curved interfaces and pores, as well as the response to electrostatic fields of these systems.

With this need in mind, we have constructed a high-throughput structured illumination second harmonic microscope to perform real-time 3D chemical imaging of surfaces and confined geometries. Non-resonant coherent second harmonic (SH) generation occurs only when non-centrosymmetric molecules, such as water, are oriented non-centrosymmetrically (*18*). This selection rule allows for probing just several monolayers at the interface. We use this property to image the orientational order of water molecules at the inner and outer surfaces of a glass micro-capillary immersed in aqueous solution. The surface chemistry of the micro-capillary is altered by changing the bulk pH of the solution. The images are converted in maps of the surface potential and surface chemical equilibrium constants. Although average surface $pK_{a,s}$ values agree with literature, a clear structural interfacial heterogeneity is apparent with $pK_{a,s}$ values ranging from 2.3 to 10.7. Dynamical structural changes induced by an external oscillating electrostatic field are also imaged on millisecond time scales and in three dimensions. The observed structural fluctuations at the interface and in the < 1 μm tip opening originate from water molecules that interact with the external electrostatic field.

The wide-field (*19-21*) structured illumination / HiLo (*22*) second harmonic microscope is sketched in Fig. 1a. To achieve the wide-field geometry the illuminating 1036 nm, 200 kHz, 168 fs pulsed beam is split into two beams with a cosine diffraction phase grating displayed on a spatial light modulator (SLM). The SLM is then imaged on the focal (sample) plane of the objectives with the aid of a relay system. The 150 μm FWHM field of view is determined by the spot size on the SLM and the overall magnification of the relay system. The spatial frequency σ of the diffraction grating determines the angle of incidence of the beams on the sample. The sample is imaged onto an EM-ICCD camera. Nonlinear polarimetry (*23*) is performed by controlling and analyzing the polarization state of the illuminating and emitted beams. A polarization state generator (PSG), comprised of a half- and a quarter-wave plate, is used. The polarization state of the emitted light is analyzed with a half-wave plate (HWP) placed in the emission path followed by a polarizing beam splitter (PBS). Further details can be found in the supplementary material (SM: M1-M2, S1-S3).

3D imaging is not possible in a wide-field illumination geometry because the beam(s) are too wide. To achieve 3D imaging and to improve the transverse spatial resolution (*24*), we use an adapted HiLo imaging procedure (*22, 25-30*), which relies on partial spatial coherence introduced by the interaction of a chirped illuminating pulse with the sample. The procedure uses Fourier filtering to reject out of focus light (SM, S1,S2). The grating provides a reference spatial frequency for the filtering procedure, using a cosine intensity pattern. With this pattern structured images $I_n$ and $I_{n\prime}$ are produced, the latter being phase shifted by π/2. The filtering is done with a uniformly illuminated image ($I_u = I_n + I_{n\prime}$), and the structured illuminated image ($I_n$), using a HiLo algorithm (*22, 26, 29, 31, 32*).

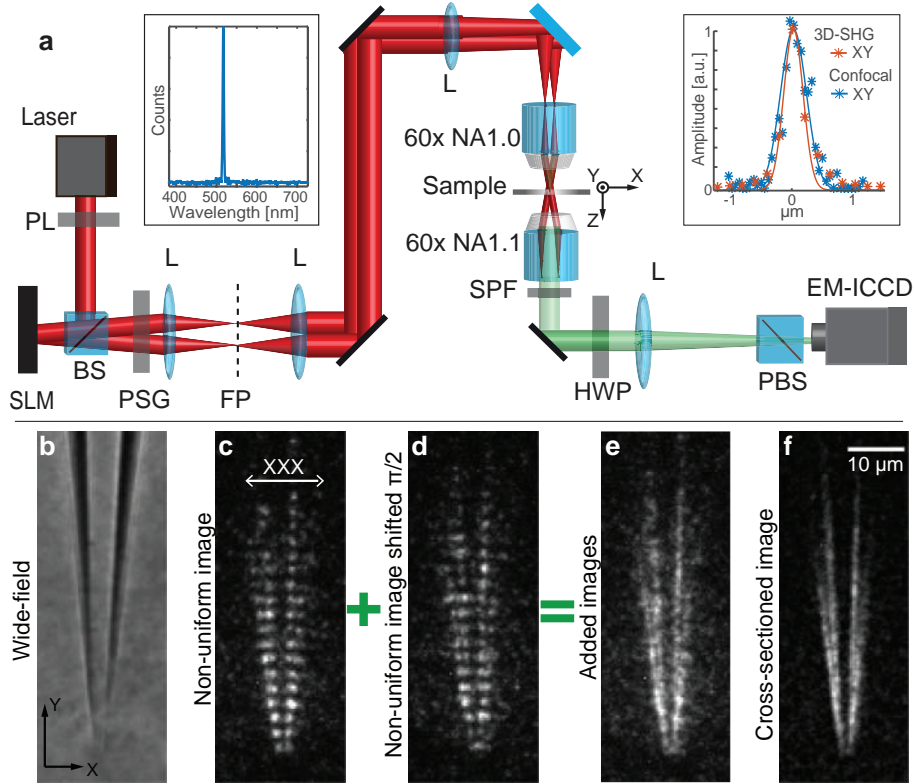

**Figure 1: Optical system and capillary imaging.** (**a**) Schematic of the 3D second harmonic microscope; SLM – spatial light modulator, PL – polarizer, BS – beam splitter, PSG – polarization state generator, L – achromatic lens, NA – numerical aperture, SPF – short pass filter, HWP – half wave plate, PBS – polarizing BS. Inset: Obtained SH spectrum from the capillary interface (left), and transverse resolution (right) of the system (red) and that of a commercial scanning multiphoton microscope (blue, Leica SP5, NA 1.2). (**b**) Phase contrast image of the micro-capillary in aqueous solution (obtained by placing a white light source in the beam path). (**c**), (**d**) SH structured image without, and with the pattern phase shifted by π/2. (**e**) Sum of (c) and (d). (**f**) HiLo image obtained after applying the HiLo algorithm (*30*) on (c) and (e) (see SM, S1,S2 for details). The SH intensity is only detectable if the polarization combination is in the XXX direction, i.e. along the surface normal.

Figures 1b-1f show phase contrast images (1b) and SH images (1c-1f) of a glass micro-capillary immersed in an aqueous solution of pH neutral ultrapure water with 10 mM NaCl. The structured SH images $I_n$ and $I_{n'}$ are shown in Figs. 1c and 1d. Figure 1e shows the uniformly illuminated image that is obtained from $I_u = I_n + I_{n'}$. Using the algorithm of Mertz and Kim (*30*) we obtain the final HiLo image of Fig. 1f (see SM: S1, S2 for details). It can be seen from comparing Figs. 1e (unprocessed, not cross-sectioned) and 1f (HiLo processed) that the background rejection results in an image with a much better contrast. The corresponding spectrum (Fig. 1a) consists of a single peak centered at 515 nm confirming that the image contrast originates from SH emission. The SH intensity is emitted in the XXX polarization combinations, i.e. with all beams polarized and analyzed in a direction perpendicular to the main symmetry axis of the micro-capillary (Fig. 1c). Light polarized along the interface does not generate SH photons. The measured HWHM transverse resolution is 188 nm (Fig. 1a) as measured by imaging 50 nm diameter BaTiO$_3$ particles (*33*) on a glass coverslip. The axial resolution can reach 520 nm, but it is different for fluorescence and SH imaging (see SM S2, Fig. S2). The imaging throughput is improved significantly compared to a scanning confocal two photon microscope (*34*) and to our previous 2D wide field SH microscope (*35*), see SM, S3.

Surface chemical changes are induced on the glass micro-capillary surface by changing the pH of the solution. SH surface images are made of three different micro-capillaries (sketched in Fig. 2a). The silica surface in contact with a solution of pH=2 is on average charge neutral, with the vast majority of the silanol groups in the protonated -SiOH state (*36, 37*). As sketched in Fig. 2b, increasing the bulk pH of the solution increases the fraction of surface SiO$^-$ groups, resulting in a predominantly negatively charged surface (with average charge densities of -2 μC/cm$^2$ under pH neutral conditions and <-17 μC/cm$^2$ at pH = 12 (*36, 37*)). Upon increasing the net surface charge by deprotonation, the interfacial electrostatic field $E_{DC}$ grows in magnitude. Interaction of this field with the dipole moment of water results in a distortion of the orientational distribution of water molecules. In absence of any external forces there is no net directionality of the water molecules (and thus no coherent SH is generated by them). In the presence of an interfacial electrostatic field the orientational distribution of water molecules changes. The dipole moment of the water molecules aligns with the electrostatic field, allowing for SH photons to be generated ((*38*), Fig. 2b). This electric-field induced SH generation (*39*) can be used to probe aqueous silica interfaces (*38, 40*). Time- and spatially averaged spectroscopic SH measurements have shown that even though the breaking of centrosymmetry occurs over a distance of only a few nm and involves only a very small fraction of all water molecules, it is enough to result in a detectable SH intensity. The spatially and temporally integrated SH intensity can be correlated to the surface potential ($\Phi_0$): $I(2\omega) \sim \left| \chi_s^{(2)} + \chi^{(3)\prime} \Phi_0 f_3 \right|^2$ (*38, 41*), which originates from:

$$I(2\omega, x, y, t) \sim I(\omega, x, y, t)^2 \left| \chi_s^{(2)}(x, y, t) + \chi^{(3)\prime} f_3 \int_0^\infty E_{DC}(x, y, z) dz \right|^2, \quad (1)$$

where $x, y, z$ represent the local surface coordinate system, with z being the surface mal, $E_{DC}(x, y, z)$ relates to the electrostatic potential change ($\Phi$) via $-\nabla \Phi(x, y, t)$, $\chi_s^{(2)}$ is the second-order susceptibility of the capillary silica/water surface, and $f_3$ is an interference term which has the value 1 for SH transmission experiments (*41*).

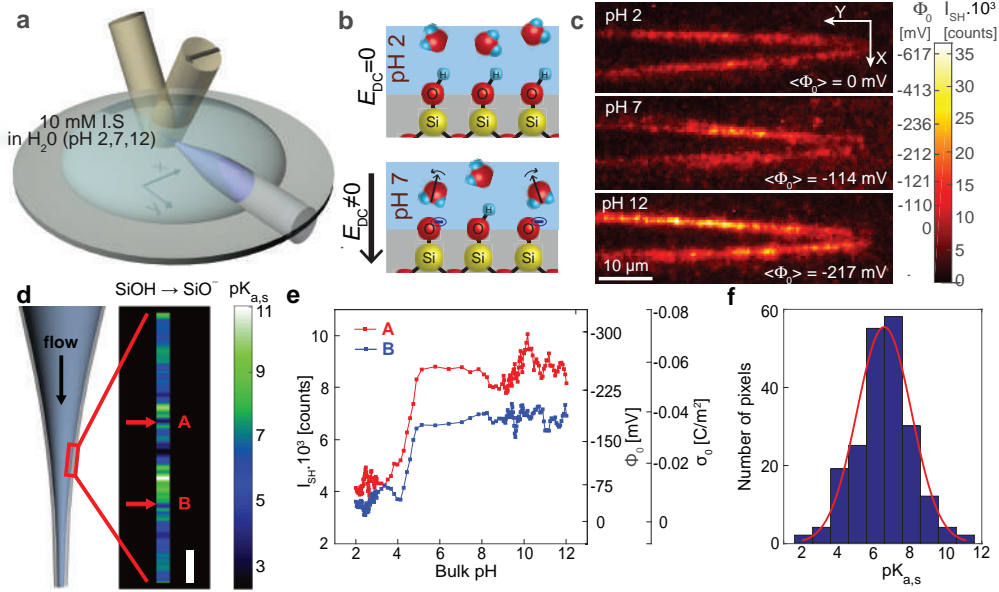

**Figure 2: Imaging of surface chemistry and potential.** (**a**) Illustration of the experiment. (**b**) Illustration of the silica surface under acidic and pH neutral conditions. When the surface is negatively charged, water molecules will be oriented by the electrostatic surface field, giving rise to an increase in the SH intensity. (**c**) SH cross-sectioned images (2 μm thick) using the XXX polarization combination for different pH values of the electrolyte solution, recorded with 5 s acquisition time per image. The color bar reports on the intensity as well as the calculated change in the surface potential ($\Phi_0$). (**d**) $pK_{a,s}$ values for the deprotonation reaction of the silica/water interface along the wall of a 100 μm wide capillary, obtained by imaging the change in the surface potential while changing the pH under laminar flow from 2 →12. The pH is determined using a solution of pH indicator dyes in a separate experiment (SM, M3). Note that the values are integrated along the X direction. Scale bar: 10 μm. (**e**) The change in SH intensity as a function of pH for two different points in (d), $pK_{a,s}$(A)=3.8 and $pK_{a,s}$(B)=5.9. The extracted surface potential and computed surface charge density values are also displayed as y-axes. (**f**). Histogram of obtained $pK_{a,s}$ values, reflecting the variance in surface reactivity. The red line represents a Gaussian fit.

$\chi^{(3)\prime}$ is an effective third-order susceptibility that contains different contributions (S4, SM, Eq. S4) of which the reorientation of water molecules in the electric double layer is the biggest one ((*41*), here ~99%, SM S4). This response can be used to determine changes in the surface potential ($\Phi_0$).

When imaging three different systems consisting of a solution with a different bulk pH value (pH = 2, neutral, and 12) and a micro-capillary immersed in it, we observe very different SH responses as shown in Fig. 2c. Note that the ionic strength is kept constant at 10 mM (by adding NaCl). It can be seen that the average intensity increases in the order pH(2) < pH(neutral) < pH(12), in agreement with expectations from spectroscopic measurements (*38, 42-45*). Using Eq. 1 we convert the SH intensity in the images of Fig. 2c to a change in the surface potential $\Phi_0$. Although it is not possible to determine absolute values for the surface potential in this experiment (*41*), assuming that $\langle\Phi_0\rangle$ =0 at pH=2 and that the average charge density under pH neutral condition is -2 μC/cm² (*36, 37*) we can construct a scale for $\Phi_0$ using the Gouy-Chapman-Stern model to provide reference values (*37*). The retrieved values are given by the color bar in Fig. 2c. The average values of $\Phi_0$ agree well with literature values (*37, 42-44*). The variation in the spatial distribution of intensities/surface potentials values is, however, significant reaching values of ~3⟨$\Phi_0$⟩. This variation reflects the diverse values reported in literature (*37, 42-44*) and indicates that the surface is structurally not uniform, for example in terms of the distribution of charged SiO⁻ or the nanoscopic surface geometry imposed by the production procedure of the capillary (see SM S5 for more analysis). Indeed, experimental reports have been made of mono-, bi-, or trimodal deprotonation behavior, with surface pK$_a$ values of 6.8 (*46*), 4.8, 8.5 (*38*), or 3.8, 5.2, and ~9, as well as a dependence on the pH history of the sample (*44*). Computations show that the local chemical structure and hydrogen bonding environment, as well as strains or defects in the surface structures all contribute (*47, 48*). Given the importance of silica as one of the most abundant minerals on planet earth, it makes sense to extend these studies beyond the time and spatially averaged mean field behavior to explain this large discrepancy. We therefore determine the $pK_{a,s}$ values for the $\equiv SiOH_{(S)} + H_2O \rightleftarrows \equiv SiO^-_{(S)} + H_3O^+_{(S)}$ reaction by flowing a pH=12 solution over an on average charge-neutral capillary (Fig. 2d-2f), using a shear rate of $8.5 \cdot 10^3$ s⁻¹. We then image surface potential changes every 250 ms. The pH of the solution was determined with a pH indicator dye solution (SM, M2). Knowing the bulk pH value inside the capillary we can determine the $pK_{a,s}$ value of every pixel using:

$$K_{a,s} = [H_3O^+_{(b)}]\frac{(N_{\equiv SiO^-_{(s)}})}{(N_{\equiv SiOH_{(S)}})} e^{-\frac{e\Phi_0}{k_b T}} \quad \text{and} \quad pK_{a,s} = -\log[K_{a,s}] \qquad (2)$$

with $\frac{(N_{\equiv SiO^-_{(s)}})}{(N_{\equiv SiOH_{(S)}})}$ the ratio of deprotonated over protonated groups, which can be derived from $\Phi_0$ (using Gouy-Chapman-Stern theory (*49*), SM, S5). Figure 2d shows the distribution of $pK_{a,s}$ values for the deprotonation reaction, integrated along the X direction. Figure 2e displays the surface potential and derived corresponding surface charge density for two pixels as a function of pH (with $pK_{a,s}(A)$=3.8 and $pK_{a,s}(B)$=5.9). Figure 2f displays histograms of the obtained $pK_{a,s}$ values with an average of 6.7. The chemical reactivity varies spatially from $2.3 < pK_{a,s} < 10.7$ (following a Gaussian distribution). The average value agrees with Ref. (*46*) but the significant variation in chemical reactivity across the surface is remarkable. It explains why different studies (even employing the same methods (*38, 44*)) find different average $pK_{a,s}$ values.

To further image dynamical changes at interfaces and in confinement, two Ag/AgCl electrodes are placed in a pH neutral solution that also contains the micro-capillary and 10 mM NaCl as sketched in Fig. 3a. The electrode outside the capillary is grounded and the external electrostatic potential difference ΔV is varied sinusoidally on the inner electrode from 10 V to -10 V with a frequency of 0.05 Hz (~1 V/s). The spacing between the electrodes is 1 cm, and we image the 60 µm long edge that has an opening with a < 1 µm radius. By solving numerically the Navier-Stokes, Nernst-Planck, and Poisson equations, using a COMSOL routine, we compute the resulting average changes in the electrostatic field ($E_{DC}$) as well as the induced flow in the aqueous solution ((*50*), SM: S8). An example of the resulting electrostatic field lines is sketched in Fig. 3b. The electrostatic field strength is high (>$10^6$ V/m, Fig. S9) in the interfacial region and in the tip of the micro-capillary. Figure 3c shows the computed average ionic charge density distribution close to the surface for different values of ΔV (using an initial surface charge density of -2 µC/cm$^2$ (*36, 37*), SM: S8). As $\Delta V$ is cycled from 10 V to -10 V the electrostatic field at the surface (z=0) of the micro-capillary changes as the distribution of Na$^+$ and Cl$^-$ ions in the electric double layer changes (Fig. S9).

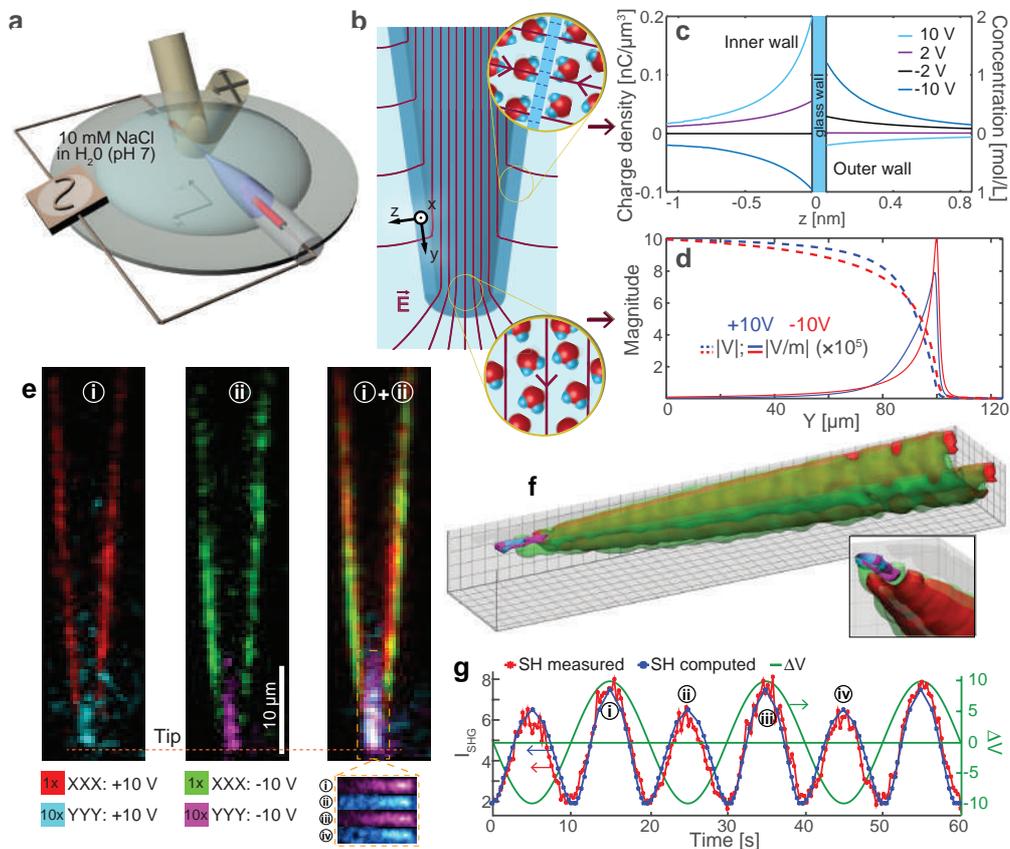

**Figure 3: Dynamic imaging of surface chemical changes.** (**a**) Illustration of the experiment. (**b**) Illustration of the external field lines (ΔV > 0). (**c**) Computed ionic charge density in the aqueous solution adjacent to the interface (details SM: S8, Fig. S7). (**d**) Electrostatic field and potential drop for ΔV=10 V and ΔV=-10 V along the symmetry (Y) axis of the capillary. (**e**) Snapshots of movie M1, recorded with 250 ms acquisition time for different values of applied potential (10 V, i; -10 V, ii; composite of 5 frames each, i+ii). All images were normalized with respect to the illuminating beam profile. (**f**) 3D render of the 10 V and -10 V channels (45 stacks per 4° rotation around the Y axis). (**g**) ΔV (green), the spatially integrated SH intensity (XXX polarization, red), and the numerical computation (blue, SM: S8).

Water molecules that are dynamically oriented by the interfacial DC field can be probed with the XXX polarization combination, as in Figure 2. Another place where the electrostatic field reaches similarly high values is inside the < 1 μm radius opening of the micro-capillary (*50*). As sketched in Fig. 3d, here, $E_{DC}$ oscillates between $0.75·10^6$ V/m and $-1.15·10^6$ V/m, which corresponds to a fraction of $5·10^{-4}$ oriented water molecules that are fully aligned by it (Fig. S4a), or, equivalently, ~1 million aligned water molecules per pixel. These aligned water molecules can be probed with the YYY polarization combination, that is, with all beams polarized and analyzed along the main symmetry axis of the capillary.

To image the change in the orientational order of water molecules in both regions as a function of time and applied potential, we recorded SH images with 250 ms integration time in different Z planes and different polarization combinations (XXX to probe the interfacial water and YYY to probe the water in the tip). Movie M1 shows the temporal evolution of the SH intensity. Figure 3e shows 250 ms snapshot cross-sections for $\Delta V$=10 V (i) and -10 V (ii) and their overlap (averaging 5 frames) taken at 10 V and -10 V. The 250 ms acquisition time is enough to resolve the interfacial water as well as the ~$10^6$ oriented water molecules in the tip. Figure 3f shows a 3D rendering constructed from 45 image stacks, with each image taken at 4° angle increments. Movie M2 shows a composite 3D rendering of the oriented water at 10 (red, cyan) and -10 V (green, purple). It can be seen (Fig. 3e/Movie M2) that there is relatively more oriented water on the inside surface for $\Delta V$>0 (also Fig. S11), which agrees with the higher electrostatic field strengths (Fig. 3c/Fig. S9). Figure 3g shows the applied potential difference $\Delta V$ (green), the spatially averaged SH intensity of the interfacial water (red, XXX polarization), and a numerical computation of the intensity using Eq. 1 (blue, see SM S9 for details). Good agreement is achieved between the mean field model and the averaged SH intensity changes. However, inspecting the surface response in the images of Fig. 3e, or movie M1, it can be seen that the intensity distribution in the images is not uniform. This difference points towards local variations in the surface structure, potential and chemical reactivity similar to those observed in Fig. 2 (see also SM, S7).

Thus, dynamical interfacial changes as well as small amounts of oriented confined water can be imaged in three dimensions and on millisecond timescales. The real-time observation of heterogeneity/transient structures in the surface charge distribution, chemical reactivity, and/or the spatial variations in flow are evidence that surface chemical processes heavily depend on local structural/chemical fluctuations on sub-micron length scales. Being able to non-invasively observe such variations optically, label-free, and in-situ in a confined geometry that is difficult to access will further our understanding of surface chemistry beyond that of idealized interfaces and the mean field and static level. This is especially necessary for the development of micro- and nanotechnology that critically relies on local structural and dy-namical processes (*11*), but also for geochemistry, catalysis, electrochemistry (such as in fuel cells) (*5, 11, 14, 17*), and processes using electrokinetic phenomena, in micro and nanofluidic environments (*51*). Being able to image ~$10^6$ oriented water molecules per pixel, and knowing that electrostatic fields of the order of $10^9$ V/m are predicted in pores that are ~1-10 nm (*13*) wide, we estimate that water in such nanopores can easily be imaged on sub-second time scales. The possibility to image the orientational order of water in confinement has great merit as the water structure can be used to non-invasively report on the structure of other macromolecules, such as lipids, DNA and proteins, or on potential gradients. The current interest in nano-confinement (*52, 53*) and the technological development and use of artificial (*12, 14, 54*) or natural (*13*) nanopores and capillaries for synthesis (*55*), chemical separation (*11*), and even power generation (*15, 16*) will benefit from it.

**Supplementary Material**

Materials and Methods

Supplementary Text

Figures S1 to S12

Movies M1, M2

References

Supplementary Materials for

**Optical Imaging of Surface Chemistry and Dynamics in Confinement**

Carlos Macias-Romero, Igor Nahalka, Halil I. Okur, Sylvie Roke[*]

[*]Corresponding author. E-mail: sylvie.roke@epfl.ch

**This PDF file includes:**

Materials and Methods
        M1. 3D Second Harmonic Microscope
        M2. Cross-sectioned images
        M3. Imaging of surface chemistry and surface dynamics

Supplementary Text
        S1. HiLo Fourier filtering procedure
        S2. Using HiLo routines for SH imaging
        S3. Imaging throughput
        S4. The SH response
        S5. Determining the surface chemical reaction equilibrium constant
        S6. SHG probing depth estimate
        S7. Spatial and temporal heterogeneity
        S8. Numerical model
        S9. Modeling the SH data of Figure 3

Figures
        S1: Illustration of the imaging process
        S2: Axial scans and resolution
        S3: Comparison of imaging throughput and signal-to-noise ratio.
        S4: Probability and SH probing depth
        S5 Spatial and temporal heterogeneity
        S6: COMSOL simulation
        S7: Simulated volumetric charge density in the EDL
        S8: Computed electrostatic field strength and ionic flow.
        S9: Derived electrostatic field at the surface
        S10: Simulated integrated SH intensity
        S11: Comparison of the simulated and measured SH intensity.
        S12: Simulated and measured SH intensity

Movies
        M1: Temporal variations in the SH intensity from the micro-capillary when a sinusoidal voltage difference is applied.
        M2: A composite 3D rendering of the oriented water at 10 V (red, cyan) and -10 V (green, purple).

References

**Materials and Methods**

**M1. 3D Second Harmonic Microscope**

The light source is an amplified laser system (Pharos SP-1.5, LightConversion) delivering 168 fs laser pulses centered around 1036 nm (+/- 6 nm) with a beam diameter of 3.7 mm and a maximum output power of 6 W at variable repetition rates between 1 kHz -1 MHz. The laser system is operated at 200 kHz, and 1.10 W and delivers 110 mW on the sample (3.6 mJ/cm$^2$). The field of view has a FWHM diameter of 150 µm. All microscope mirrors are protected silver mirrors (Thorlabs, PF10-03-P01). The (achromatic) lenses and other optical elements on the illumination path are near-infrared antireflection coated (Thorlabs, B), whereas the optical elements on the detection path are antireflection coated for the visible region (Thorlabs, A). The water immersion objective lenses have the following specifications: The bottom objective lens (Olympus, LUMFLN 60XW) has a 60x, NA 1.1, 1.5 mm working distance, and a collar for variable coverslip correction. The top objective lens (Olympus, LUMPFLN 60XW) has a 60x, NA 1.0, 2 mm working distance, and no collar for variable coverslip correction. The tube lenses are provided by the manufacturer of the objective lenses (Olympus, U-TLU). The SLM (Holoeye, Pluto-NIR-015) is a phase only device coated for near-infrared wavelengths, and it can withstand fluences of < 20 µJ/cm² (at 1036 nm, 168 fs).

The sample is mounted on an XYZ translation stage (Asi Imaging, PZ-2000). The XY-axes are controlled by actuators and have 5 cm travel range, whereas the Z-axis consists of a piezoelectric stage with 300 µm travel range. The objective lenses are mounted on single axis actuator stages (Asi Imaging, LS-200). All the stages are mounted on a modular system (Asi Imaging, RAMM). The imaging camera is a back-illuminated electron-multiplied and intensified CCD camera with 512 x 512 pixels (PI-MAX4: 512EM-HBf P46 GEN III).

The transverse HWHM resolution (188 nm) was determined by imaging 50 nm diameter BaTiO$_3$ particles on a glass coverslip (see Ref. (*1*) for details on the particles and imaging procedure). The maximum HMHM axial resolution, achievable in fluorescence or TPF imaging was determined by scanning a 300 nm thick aqueous solution of 9-diethylamino-5-benzo[α] pheno-xazinone (Nile Red) through the focus. Using a spatial frequency of σ=0.83 µm$^{-1}$ we obtain an axial resolution of 520 nm. Compared to our previous 2D wide-field microscope (*2*) the resolution in the transverse direction is improved by a factor >2, and the Z-sectioning ability is new. Compared to an in-house Leica TSC SP5 commercial scanning confocal multiphoton microscope, used to measure the same spot on the same sample operated with the same pulse duration (described in (*2*)), wavelength and NA, the transverse resolution is improved somewhat (by a factor of ~1.4). Compared to spatiotemporal imaging as performed in Refs. (*3-6*) the transverse resolution is comparable (with also an improvement of a factor

~2). This difference is discussed in more detail in section S2, where we also plot a graph of the axial resolution of our system as a function of spatial frequency and discuss the result in more detail.

**M2. Cross-sectioned images**

The SH images demonstrating the HiLo procedure shown in Fig. 1 were obtained with an integration time of 5 seconds each. The micro-capillary was mounted on the microscope on a glass coverslip (#1) and immersed in a pH neutral 10 mM NaCl solution. The gain of the camera was set to 7000 x in automatic mode (i.e. setting the optimum gain between the intensifier and electron multiplier internally). The images of the stack were taken 500 nm apart by moving the Z-piezo stage. The polarization of the illuminating beams is directed perpendicular to the walls of the micro- capillary. The polarization is analyzed along the same direction. A program written in LabView controls all stages and the image acquisition. The raw data is processed with a HiLo algorithm (see Sec. S1 below) written in Matlab. The coefficient $\mu$ in the HiLo algorithm is set to 6. The spatial frequency ($\sigma$) of the illuminating pattern is $\sigma=0.64$ $\mu m^{-1}$, which corresponds to an angle of 41.5 degrees between the beams. The power on the sample is 110 mW. The images are flat-fielded by the transverse beam profile using a uniform fluorescence target and the pixel-response of the CCD camera.

**M3. Imaging of surface chemistry and surface dynamics**

NaCl (99.999%, 38979 Sigma-Aldrich), NaOH (99.99%, 306568 Sigma-Aldrich), and HCl (99.99%, 306568 Sigma-Aldrich) were used as received without further purification. All samples were made by dissolving the electrolytes in degassed 18.2 M$\Omega$·cm ultrapure water that was obtained from a Milli-Q UF plus instrument (Millipore, Inc.). The electrolyte solution with pH 2 was obtained by adding 10 mM of HCl to ultrapure water. The pH neutral solution was obtained by adding 10 mM of NaCl to ultrapure water. The solution with pH 12 was obtained by adding 10 mM of NaOH to ultrapure water. The micro-capillary (MGM-1x, FIVEphoton biochemical) was then filled with the electrolyte solution, placed on a glass coverslip (#1), mounted on the microscope stage and immersed in the electrolyte solution with the same pH; the pH is the same on the inside and the outside (Fig. 2 and 3). For Fig. 2c each image was obtained with an acquisition time of 5 s and HiLo processed. The illuminating power at the sample was 110 mW. The gain of the camera was set to 7000 x in automatic mode.

For the flow experiments (Fig. 2d) a capillary was connected to silicon tubes attached to a syringe pump (Harvard Apparatus PHD 2000). The diameter of the capillary was 100 $\mu$m were the images were taken, and the diameter at the tip was 30 $\mu$m to enable a steady flow. With a flow rate of 50 $\mu$L/min a shear rate of $8.5 \cdot 10^3$ $s^{-1}$ was achieved, resulting in a laminar flow. Starting with a pH neutral solution with 10 mM NaCl for obtaining a reference, the solution was changed to pH=2 that was flown at a speed of 0.5 mL/min for 5 minutes to completely protonate all silica groups. SH images were recorded with 250 ms acquisition time, using σ=1.04 µm$^{-1}$. The data was not HiLo processed and the SH signal was integrated in the X direction to improve the signal to noise ratio. The illuminating power at the sample was 110 mW. The gain of the camera was set to 7000 x in automatic mode. The bulk pH of the solution as a function of time was determined by repeating the flow experiment on the same spot of the same capillary inside a UV/Vis spectrometer (Varian Cary 50 Bio) adding a few drops (250 µL) of pH indicator dye to the solution, as specified (Honeywell Fluka 31282, Universal indicator solution).

For the dynamics experiments (Fig. 3) the micro-capillary was filled with and immersed in pH neutral degassed ultrapure water containing 10 mM of NaCl. An Ag/AgCl wire electrode (36 AWG, WPI) was placed inside the capillary and an AG/AgCl pellet electrode (EP05, WPI) was placed in the liquid facing the small opening of the capillary (Fig. 3a). The micro-capillary and electrodes were placed on a glass coverslip (#1), and mounted on the microscope stage. The electrode placed outside the micro-capillary was held at 0 V and the electrode inside the micro-capillary was set to $\Delta V = V_0 \sin \upsilon t$ with $\upsilon = 0.05$ Hz and $V_0$=10 V using a function generator (Hewlett-Packard 3312A). The SH images were recorded with 250 ms integration time and HiLo processed. The illuminating power at the sample was 110 mW. The 3D render was produced with 1 s images, by averaging 4 250 ms frames and then rotating the capillary by 4º, resulting in a stack of 45 images (recorded in 45 s) The gain of the camera was set to 7000 x in automatic mode.

# Supplementary Text

**S1. HiLo Fourier filtering procedure**

The 3D imaging is achieved by implementing an adapted HiLo imaging procedure (*7-13*). In this procedure a first image ($I_n$) is recorded with the sample being illuminated with a patterned intensity distribution (typically with a single spatial frequency). A second image ($I_u$) is then obtained but with the sample illuminated with a uniform intensity distribution. The Z-sectioned image is created from these two images with the aid of the algorithm of Ref. (*12*) in which low ($I_{Lo}$) and a high ($I_{Hi}$) spatial frequency content images are obtained from:

$$I_{Lo} = F_2\{|F_1\{I_u - I_n\}|\}, \tag{S1}$$

and

$$I_{Hi} = F_2^*\{I_u\}. \tag{S2}$$

Here $F_1\{\ \}$ denotes a high-pass filter operator intended to center $I_u - I_n$ around zero, $F_2\{\ \}$ denotes a low-pass Gaussian filter operator and $F_2^*\{\ \}$ is its conjugate (high-pass) operator. The cut-off frequency of $F_2$ is typically set at half of the spatial frequency of the illumination pattern. The HiLo image (of Fig. 1e) is then given by:

$$I_{HiLo} = I_{Hi} + \mu I_{Lo},$$

where $\mu$ is a parameter chosen empirically to aid the transition between low and high frequencies (*12*). The image $I_n$ is recorded with a cosine phase grating and $I_{n'}$ is recorded with the same grating shifted by a quarter of a period (π/2). Adding them results in the uniform image ($I_u$), as $cos^2(\sigma x) + cos^2\left(\sigma x + \frac{\pi}{2}\right) = 1$. This is different from HiLo procedures reported in the literature, where the uniform and structured images are obtained by changing the angle of incidence of the light or by randomizing the wave front (*6, 8, 13, 14*). These approaches cannot be used with pulsed illumination, however.

## S2. Using HiLo routines for SH imaging

HiLo procedures are based on Fourier filtering and typically applied for fluorescence imaging. For this type of imaging the emitted light is spatially incoherent. This incoherent nature of the fluorescent signal ensures that outside of the focus, the interference fringes decay rapidly in the axial (Z) direction. This process is illustrated in Fig. S1. The rapid decay is essential to obtain cross-sectioned images with maximum axial resolution.

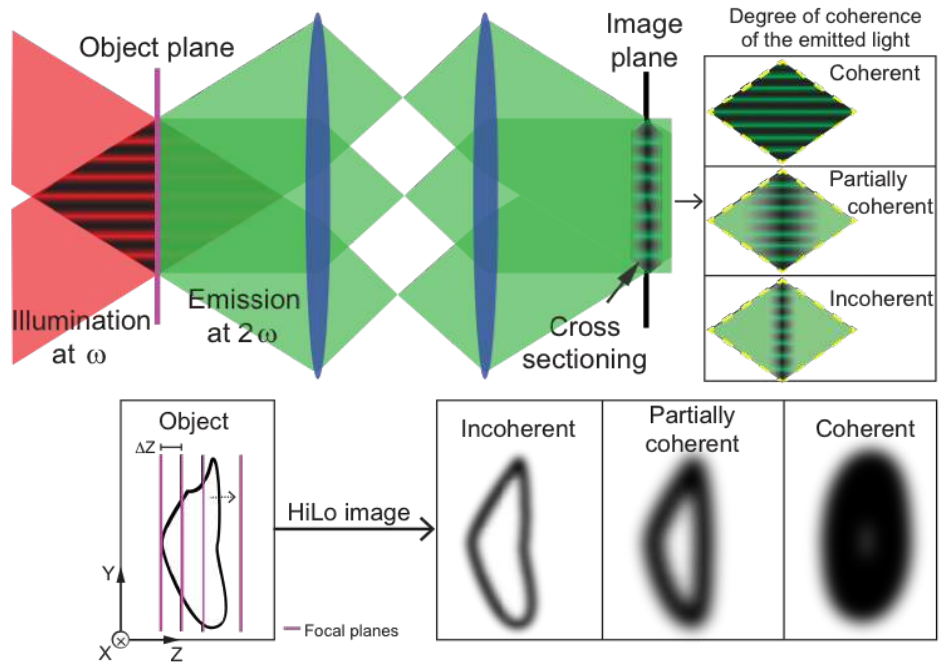

**Figure S1: Illustration of the imaging process.** The axial resolution is expected to vary with the amount of coherence in the emitted light. For incoherent emission a good axial resolution is achievable. For fully coherent emission, no axial resolution is achievable. For partially coherent light an improved axial resolution is expected.

If the emitted light is spatially coherent, interference fringes will persist and modify the axial resolution. Figure S1 also illustrates what will happen for the case of fully coherent emission. Here, an interference pattern is created along the axial direction, which will not decay, and thus no cross-sectioning capabilities are achieved. If the emitted light is only partially coherent, we expect that there will be some intermediate scenario, with some amount of spatial coherence present in the emitted light. In this case, there are fringes, and they do decay but over some distance (illustrated in the middle panel of Fig. S1). In this case HiLo imaging can still be performed but with a lower axial resolution than for fluorescence imaging.

In order to verify this hypothesis we scanned a monolayer of hexagonal boron nitride (hBN) that was deposited on a glass surface through the focus and measured the fringe pattern. The hBN 2D films are single crystalline (*15*) and with non-resonant SHG the emission process does not alter the coherence of the light. Since the thickness is only one monolayer, it can be used to map the axial point spread function (PSF) of the system.

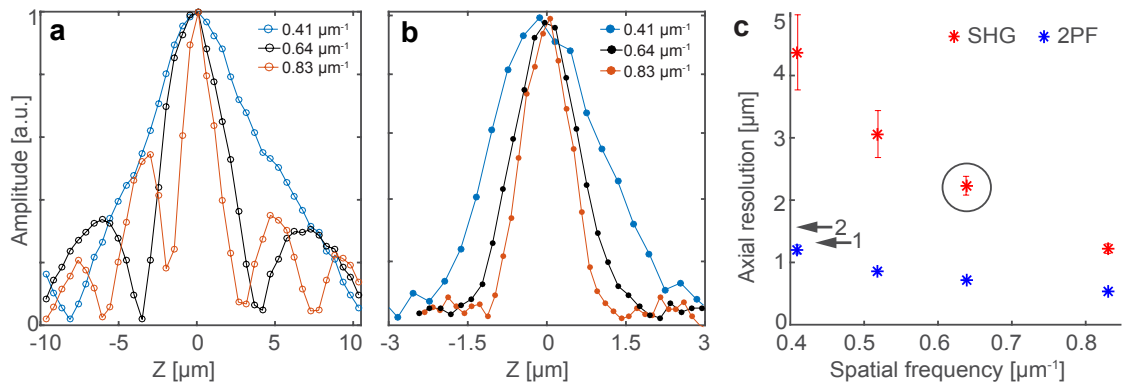

**Figure S2: Axial scans and resolution. a.** Interference fringes along the axial (Z) direction for SH imaging, obtained by scanning a hBN monolayer along the axial direction. **b.** Interference fringes along the axial (Z) direction for TPF imaging, obtained by scanning a 300 nm thick solution of 9-diethylamino-5-benzo[α] pheno-xazinone (Nile Red) along the axial direction. **c.** Axial resolution of the 3D SH microscope as a function of the spatial frequency of the illuminating pattern (red) and for TPF (blue). The arrows indicates the axial resolution measured with the scanning confocal system under the same illumination and detection conditions (*1*), and a measurement by Choi et al. (*5*).

Figure S2a displays scans of the hBN monolayer through the axial direction for four different spatial frequencies. For all four spatial frequencies there is a peak around Z=0 that is significantly more intense than the side lobes. As we expect that the hBN response function is fully coherent, this indicates that the illuminating light is not fully coherent, a common feature of amplified femtosecond laser pulses (*16*). As a consequence the PSF is not a pulse train consisting of peaks with equal intensity, but consists of a main peak with side lobes. The axial resolution is determined by the HWHM of the main peak.

To understand the changes in axial resolution due to the different degrees of coherence, consider imaging a plane that is illuminated with a single wavelength plane wave and emits light with a different wavelength and a certain degree of spatial coherence. From the theory of partial coherent imaging (*17*), we know that the light arriving in the image plane can be decomposed in a series of plane waves arriving at different angles that add in intensity plus other plane waves that add in phase (depending on the degree of coherence). The plane waves that add in phase are responsible for the presence of the inference fringes along the Z axis, and the plane waves that add in intensity are responsible for reducing the visibility of the interference fringes along the Z axis. If the plane that is imaged is illuminated with a cosinusoidal pattern, this pattern will always be present at the focus of the image plane, since it is the image of the illumination. The visibility of interference along Z depends on the presence of plane waves that add in intensity. Therefore, with a decreasing degree of coherence, the visilibity of the fringes along Z reduces, resulting in a reduction in the side lobes seen in Fig. S2a, and hence an increase in the axial resolution compared to fully coherent light. In the extreme scenario of fully incoherent light, as it is the case for two photon fluorescence (TPF), the axial distribution results in a single peak as observed in Fig. S2b. It can also be seen in both Figs. 2a (SHG from hBN) and 2b (TPF) the axial resolution depends on the spatial frequency. In general, larger frequencies lead to bigger opening angles $2\delta$ (via $\sin 2\delta = \sigma\lambda$), narrower interference fringes, and higher resolution (Fig. S2c) (*18*). The axial resolution, defined as the HWHM of the central peak, is plotted in Fig. S2c for both cases. It can be seen that best axial resolution is 520 nm, obtained with σ=0.83 µm$^{-1}$ for TPF. For SH imaging values between 1.1 and 4.3 µm are obtained using 0.41<σ<0.83 µm$^{-1}$. For the measurements of Fig. 2c and Fig. 3 we used the encircled value of σ=0.64 µm$^{-1}$. For the measurement of the chemical reactivity (Fig. 2d) σ=1.04 µm$^{-1}$ was used.

Compared to the wide field system of Ref. (*2*) that has essentially no axial resolution, the HiLo procedure introduces the possibility of Z-sectioning. In confocal scanning microscopy and spatiotemporal imaging the transverse and axial resolution is determined by the angular range of plane waves that are added in the object plane (larger angular range of waves are added with a higher NA). The axial resolution of scanning confocal imaging is indicated by arrow 1, and that of the spatiotemporal focusing with HiLo processing is indicated as arrow 2. When the sample is illuminated by a cosinusoidal pattern, the information concerning high spatial frequencies of the sample (which would normally be lost in confocal and spatiotemporal imaging) is shifted into the collecting numerical aperture and becomes present in the frequency space (*19, 20*). Using spatiotemporal focusing with HiLo results in an improved resolution compared to spatiotemporal focusing alone (*5*) indicated by the arrow in Fig. S2c. However, the amount of information captured outside the numerical aperture is limited by the necessity to disperse the pulse in the pupil plane of the illuminating lens.

## S3. Imaging throughput

To evaluate the contrast and signal to noise ratio (SNR) of the imaging system, we imaged 50 nm BaTiO$_3$ particles with different acquisition times (following the procedure in (*2*)). We compared the images taken from the same location on the sample using different imaging systems but identical imaging parameters. Figure S3 shows the Michelson contrasts and SNRs obtained for three different systems as a function of the acquisition time. The systems being compared are the 3D SH microscope proposed here (red data), a 2D wide-field imaging system (*2*), WF 1.0, green data), and a commercial scanning multiphoton microscope (Leica TSC SP5, blue data). The SNRs are given in parenthesis and were calculated as the ratio of the contrast and its standard deviation over 20 measurements. In all three systems the same fluence was used (2.5 mJ/cm$^2$). An image with an SNR of 8-10 can be obtained with all three systems, but with significantly shorter acquisition times using the system proposed here: 50 µs (this system), 4 ms (Ref. (*2*)), and 1 s (scanning confocal). This translates into an increase in throughput of up to ~100 times with respect to the 2D wide field system of Ref. (*2*), and ~5000 times compared to the commercial scanning confocal system. As worked out previously (*2*), part of the improvement over the scanning confocal system originates from combining a medium repetition rate laser source with gated detection (*1, 21*). Another part of the improvement, which also applies to the comparison with Ref. (*2*) is the use of water immersion objectives of higher NA. Such improvements lead to a better transverse resolution, the ability to perform 3D scanning, and an improvement in contrast particularly for low acquisition times.

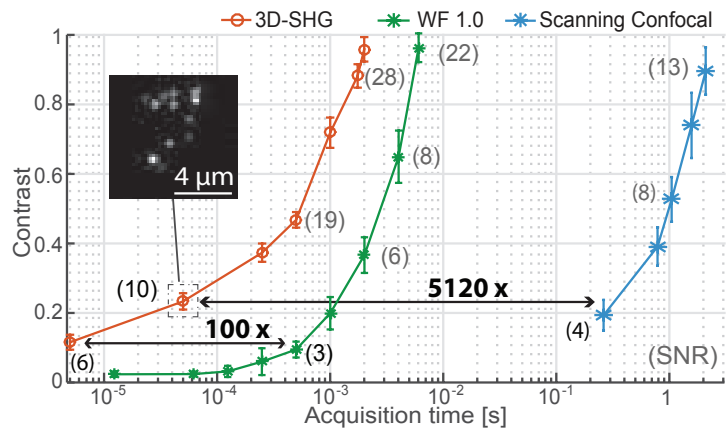

**Figure S3: Comparison of imaging throughput and signal-to-noise ratio.** Plot comparing the Michelson contrast from images of the same 50 nm BaTiO$_3$ particles taken with different acquisition times using the imaging system presented here (red), a two beam wide-field SH microscope (WF 1.0, green), and a commercial scanning multiphoton microscope (Leica SP5, blue). The number in parenthesis represents the corresponding SNRs.

## S4. The SH response

From the theory of second harmonic generation we have for the SH intensity (*22, 23*):

$$I \propto \varepsilon_0^2 \left| \chi^{(2)}{:}\, \mathbf{E}_1(\omega)\mathbf{E}_2(\omega) + \chi^{(3)\prime} \vdots\, \mathbf{E}_1(\omega)\mathbf{E}_2(\omega)\mathbf{E}_{DC} \right|^2, \tag{S3}$$

where $\mathbf{E}_{DC}$ is an electrostatic field, $\mathbf{E}_1$ and $\mathbf{E}_2$ are the electric fields of the light oscillating at a frequency $\omega$, and $\chi^{(2)}$ and $\chi^{(3)\prime}$ are the second- and effective third-order susceptibility tensors of the material in question, respectively (*23, 24*). For the effective third-order susceptibility we have (*24*):

$$\chi^{(3)\prime}_{ijkz} = g\beta^{(3)}_{ijkz} + \frac{2\mu_z}{3kT} h_1 \beta^{(2)}_{s,ijk} + \frac{2\mu_z}{3kT} h_2 \beta^{(2)}_{ijk} \tag{S4}$$

where $g$ and $h$ are constants that are determined by spatially averaging the orientational distribution of water molecules and the number of water molecules present, $\beta^{(3)}_{ijkz}$ is the third-order bulk hyperpolarizability of water, $\beta^{(2)}_{s,ijk}$ is the second-order hyperpolarizability of water at the surface (which contributes in the range $0 < z < \sim 0.3$ nm), $\beta^{(2)}_{ijk}$ is the second-order hyperpolarizability of bulk water (which contributes in the range $z > 0.3$ nm). It is commonly assumed that $\beta^{(2)}_{s,ijk} \cong \beta^{(2)}_{ijk}$. $h_1$ and $h_2$ are different since the orientational distribution of water is different. It was recently estimated that the contribution of $\beta^{(3)}_{ijkz}$ is negligible compared to the other two (0.15 %, (*25*) SM, S4).

The time varying component of the intensity normal to the walls at the surface can be written as (Eq. 1):

$$I(2\omega, x, y, t) \sim I(\omega, x, y, t)^2 \left| \chi^{(2)}_s(x, y, t) + \chi^{(3)\prime} f_3 \int_0^\infty E_{DC}(x, y, z, t) dz \right|^2 \tag{S5}$$

For non-resonant SHG we can expect that the value of $\chi^{(3)\prime}$ is the same as for other aqueous systems as it reports on the second-order hyperpolarizability and dipole moments of pure water. The value for $\chi^{(2)}_s$ will be interface dependent. To retrieve changes in the surface potential (Fig. 2) we use Eq. S5 /1 with the assumptions that at pH=2 the average charge density on the surface is zero (*26, 27*). The remaining response then originates from $\chi^{(2)}_s$. Assuming an average charge density at pH neutral conditions of -2 µC/cm$^2$ (*26, 27*), and the Gouy-Chapman-Stern model (*28*) that is commonly applied for the silica-water interface (*27*), a value of -114 mV is found for the average surface potential under pH neutral conditions and an ionic strength of 10 mM. The spatially averaged ratio $\frac{\chi^{(2)}_s}{\chi^{(3)\prime}}$ will then be -0.28. Using $\chi^{(3)\prime} = 10.3 \cdot 10^{-22}$ m$^2$/V$^2$ for water from (*29*), we obtain $\chi^{(2)}_s = -2.85 \cdot 10^{-22}$ m$^2$/V.

## S5. Determining the surface chemical reaction equilibrium constant

The surface of glass is populated with ≡SiOH and ≡SiO⁻ groups. At the glass/water interface, the population of these groups depends on the pH of the aqueous solution in contact with the surface via the following chemical equilibrium.

$$\equiv SiOH_{(s)} + H_2O \overset{K_{a,s}}{\leftrightarrows} \equiv SiO^-_{(s)} + H_3O^+_{(s)} \tag{S6}$$

where $\equiv SiOH_{(s)}$ and $\equiv SiO^-_{(s)}$ are the protonated and deprotonated surface sites, respectively. $H_3O^+_{(s)}$ is the interfacial hydronium ion concentration. $K_{a,s}$ denotes the equilibrium constant for the surface (s) chemical reaction, which can be expressed as follows:

$$K_{a,s} = \frac{[H_3O^+_{(s)}](N_{\equiv SiO^-_{(s)}})}{(N_{\equiv SiOH_{(S)}})} \tag{S7}$$

$[H_3O^+_{(s)}]$ refers to the H₃O⁺ interfacial ion concentration. $(N_{\equiv SiO^-_{(s)}})$ and $(N_{\equiv SiOH_{(S)}})$ represent the number of deprotonated and protonated groups per unit area. The H₃O⁺ ion concentration near the surface cannot be measured directly but it can be estimated if we know the bulk pH and the surface potential ($\Phi_0$) of the glass surface. The surface H₃O⁺ concentration is related to the bulk H₃O⁺ concentration by:

$$[H_3O^+_{(s)}] = [H_3O^+_{(b)}] e^{-\frac{e\Phi_0}{k_bT}} \tag{S8}$$

where $[H_3O^+_{(b)}]$ is the bulk aqueous H₃O⁺ ion concentration, $e$, $k_b$ and $T$ denote the elementary charge, the Boltzmann constant and the temperature. Combining both expressions we get (*30*):

$$K_{a,s} = [H_3O^+_{(b)}] \frac{(N_{\equiv SiO^-_{(s)}})}{(N_{\equiv SiOH_{(S)}})} e^{-\frac{e\Phi_0}{k_bT}} \tag{S9}$$

The $pK_{a,s}$ of this chemical equilibrium can be obtained via $pK_{a,s} = -\log K_{a,s}$. To find the value for $pK_{a,s}$, we need to know when 50 % of the available sites have been deprotonated (that is, when $\frac{(N_{\equiv SiO^-_{(s)}})}{(N_{\equiv SiOH_{(S)}})} = 1$).

As described above, the SH intensity scale can be converted into a surface potential scale. Knowing the values of $\Phi_0$ from our images, it is possible to compute the surface charge density ($\sigma_0$) per pixel using the Gouy-Chapman-Stern model (*28*):

$$\Phi_0 = \frac{\sigma_0 d_s}{\varepsilon_0 \varepsilon_{r'}} + \frac{2 k_B T}{e} \sinh^{-1}\left(\frac{\sigma_0}{\sqrt{8000 \, k_B T N_A c \varepsilon_0 \varepsilon_r}}\right) \tag{S10}$$

with $\varepsilon_0$, $\varepsilon_r'$, $\varepsilon_r$, $d_s$, $N_A$, $c$ the permittivity of vacuum, the dielectric constant in the Stern layer, the dielectric constant of water, the Stern layer thickness, Avogadro's number, and the ionic strength in the solution in mol/L. Taking the Stern layer thickness at $d_s = 0.91$ nm, and the dielectric constant in the Stern layer as $\varepsilon_r' = 43$ from Ref. (*31*), we can convert the values of $\Phi_0$ into values of $\sigma_0$. Knowing $\Phi_0$ and $\sigma_0$ as a function of time and bulk pH, we can find the point $\sigma_p$ were $\frac{(N_{\equiv SiO^-_{(S)}})}{(N_{\equiv SiOH_{(S)}})} = 1$ as $\sigma_p = \frac{1}{2}(\sigma_{pH=12} - \sigma_{pH=2}) + \sigma_{pH=2} = \frac{1}{2}\sigma_{pH=12} + \frac{1}{2}\sigma_{pH=2}$. At this pH we obtain the chemical surface reaction constants as:

$$K_{a,s} = [H_3O^+_{(b)}(\sigma_p)]e^{-\frac{e\Phi_0(\sigma_p)}{k_b T}} \text{ and } pK_{a,s} = -\log[K_{a,s}]. \tag{S11}$$

## S6. SHG probing depth estimate

Having determined that DC aligned water molecules are the major source of the SH signal, we estimate the probing depth of the SH experiment. To compute the probability (p) to find a water molecule that is aligned by the surface DC field we can use the Boltzmann distribution (*32*):

$$p = \frac{\int_0^\pi \cos\theta \, e^{-\frac{\mu E}{kT}} \sin\theta \, d\theta}{\int_0^\pi e^{-\frac{\mu E}{kT}} \sin\theta \, d\theta} \tag{S12}$$

Where $\mu$ is the dipole moment of water, $E$ the magnitude of electrostatic surface field, $k$ the Boltzman constant, $T$ the temperature, and $\theta$ the polar angle if the molecular axis with respect to the surface normal.

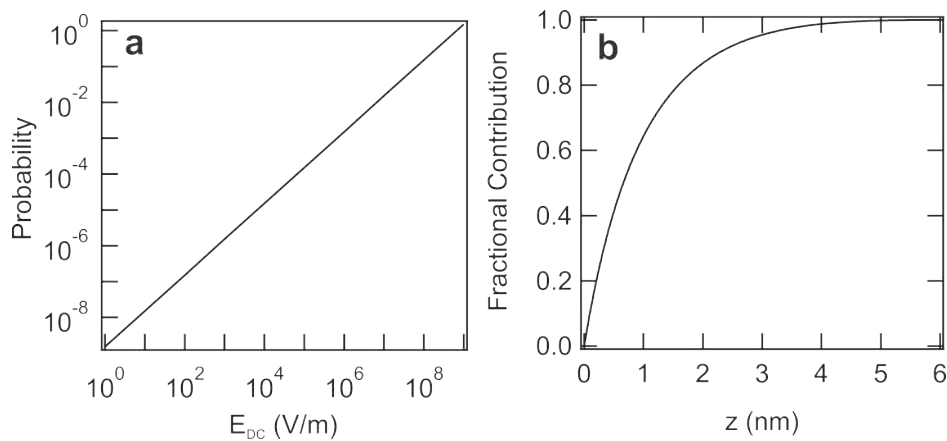

**Figure S4: Probability and SH probing depth. a.** Probability to find a water molecule fully aligned with an applied external electrostatic field. **b.** Integrated squared probability as a function of distance away from the surface normal for an idealized glass/water interface.

Fig. S4a shows the computed probability as a function of $E$. Using the electrostatic field of Fig. S8 we can compute the integrated value of the square of the probability, which estimates the z dependent contribution to the observed intensity. The result is shown in Fig. S4b. It shows that for the present conditions 64 % of the SH intensity is built up within the first 1 nm. Note that this number depends on the ionic strength.

## S7. Spatial and temporal heterogeneity

The spatial heterogeneities in the images of Figure 2c can be estimated by averaging the intensity values within squares of different pixel areas in the image and calculating the standard deviation of these larger squares throughout the mosaicked image. The standard deviation as a function of the size of the integration area is plotted in Fig. S5a. It can be seen the standard deviation becomes a factor of 1.9 smaller when the integrated area is increased from 0.16 µm$^2$ to ~20 µm$^2$, where it levels off. The transition occurs at ~ 10 µm$^2$. The spatial extend of the heterogeneity thus comprises ~125 pixels. This value of ~10 µm$^2$ suggests that the spatial heterogeneities observed here are connected to the surface roughness, which typically occurs over a length scale on the order of a micron and likely originates from the production process of the micro-capillary. Applying an external potential difference (Fig. 3) results in an initially ~20 % larger spread in the computed surface potential values. This suggests that the spatial heterogeneity is not only connected to the surface roughness but also to the distribution of ions in the electric double layer, as the external bias modifies this distribution

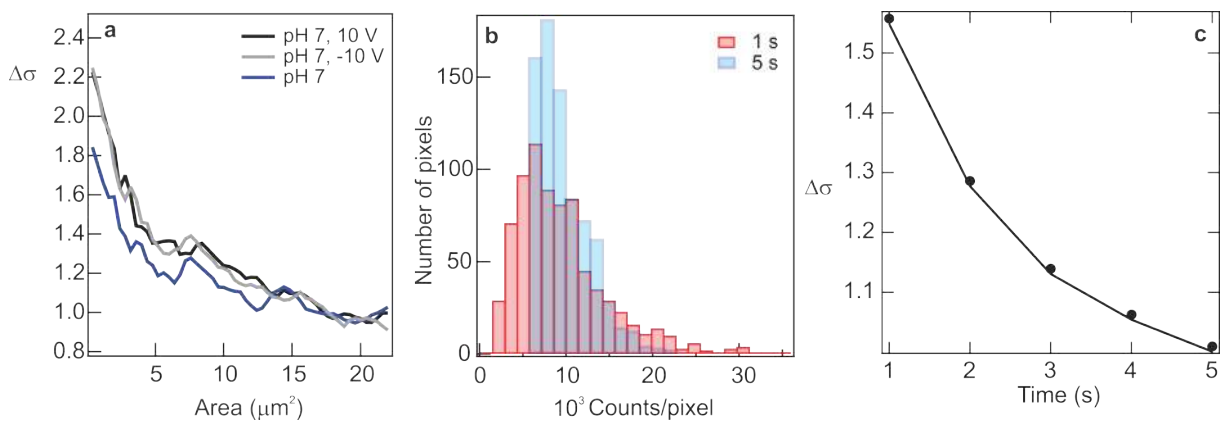

**Figure S5: Spatial and temporal heterogeneity**. **a.** Standard deviation in the intensity as a function of integration area (pH neutral, Fig. 2c and Fig. 3). **b.** Histogram of counts/pixel for a 1 s and a 5 s integration time (pH neutral, Fig. 2c). **c.** Change in the standard deviation as a function of integration time (Fig. 2c, pH neutral). The data in panels a and c is normalized to the steady state standard deviation.

To investigate the temporal behavior of the observed heterogeneity in Fig. 2c we compute the standard deviations of the intensity / pixel for the pH neutral case. Fig. S5b shows two examples of different distributions with 1 s and 5 s integration times. Fig. S5c shows the standard deviation divided by the steady state standard deviation as a function of time. It can be seen that taking snapshots of 1 s instead of 5 s results in a bigger distribution of surface potential values. This time scale of a few seconds is much longer than the timescale of diffusion in the electric double layer from where the SH intensity originates (see Fig. S8b, it is ~0.9 mm/s at a depth of 1 nm).

**Interpretation**

It has been shown that silica in contact with water undergoes several chemical reactions (*33*), of which the dissolution of silica and the deprotonation of surface –OH groups are most important (*34*). These reactions occur at different speeds, depending on the local pH and surface structure and determine the charge on the surface. The speed of these reactions (investigated on a fused silica crystal surface) depends on the ionic strength, the pH of the solution, and the flow rate of the liquid (*34, 35*).

It is clear from the experiments shown here that the structure of the capillary is heterogeneous on a micron length scale, with patches that have vastly different surface potentials, and as a consequence, different surface chemical reaction equilibria constants for the deprotonation reaction (with an average $pK_{a,s}$ of ~6.7 and limiting values of 2.3 and 10.7). This shows there is a complex surface charge distribution varying across the interface that drives an ionic flux (of all ionic species) along the surface plane. This ionic flux in turn influences the chemical conversion of the silica groups (as evidenced by the effect on the bias on the spread in surface potential values in Fig. S5a). The establishment of the local chemical reaction equilibria (and their cross-talk) occurs on a much slower time scale than the diffusion. Areas with a high reactivity will convert more quickly into other species, which explains why the distribution in Figs. S5b/c narrows.

## S8. Numerical model

**Geometry, equations and boundary conditions**

A glass micro-capillary is submerged in a water solution of 10 mM NaCl. A grounded electrode (with V = 0 V) is placed in the reservoir outside the micro-capillary, and another electrode (V=$V_0$) is placed inside. The glass of the micro-capillary is charged with -2 µC/cm$^2$ and it has a relative permittivity of 4.2. The potential of the inside electrode is then varied from -10 to 10 V with respect to the counter electrode. The simulation consists in solving the Poisson, Nernst-Planck, and Stokes (PNPS) equations for the geometry shown in Fig. S6. The geometry is rotationally symmetric with the red line being the symmetry axis. The nanometer length scale of the simulated system is smaller than the real system but captures the essence of the physical phenomena while keeping the computational time reasonable. The PNPS equations relate the ion flux of species $i$ (**J**$_i$) to the velocity field **u** of the liquid and the electrostatic potential ($\phi(r)$) (S12), the electrostatic potential to the charge density (S13), the velocity field of the liquid to the potential for a given charge density (S14), and the charge density (S15) are written as (*36*):

$$\boldsymbol{J}_i = -D_i \nabla c_i - \frac{D_i}{RT} z_i F c_i \nabla \phi(\boldsymbol{r}) + c_i \boldsymbol{u} \tag{S12}$$

$$\nabla^2 \phi(\boldsymbol{r}) = -\frac{\rho(\boldsymbol{r})}{\varepsilon_0 \varepsilon_r} \tag{S13}$$

$$\mu \nabla^2 \boldsymbol{u} = \rho_e(\boldsymbol{r}) \nabla \phi(\boldsymbol{r}) + \nabla P \tag{S14}$$

$$\rho_e(\boldsymbol{r}) = F \sum_i z_i c_i(\boldsymbol{r}) \tag{S15}$$

Here $D_i$, $z_i$ and $c_i$ are diffusivity, valence, and concentration of the $i$-th species, respectively. $F$ and $R$ are the Faraday and gas constants. $T$ is the temperature, $P$ is the pressure of the liquid, $\varepsilon_0$ is the vacuum permittivity, $\varepsilon_r$ is the relative permittivity, and µ is dynamic viscosity. Bold typeface denotes vector form. The boundary conditions and initial input parameters are summarized in Fig. S6. The equations are solved for the electric potential, the velocity field and pressure in the liquid exerted by the electrostatic fields, the concentration distribution of each of the dissolved species, and consequently the charge density everywhere in the domain.

The model is solved in a similar way to the two-step approach by Laohakunakorn *et al.*(*36*), but with both steps in 2D (with rotation symmetry), using the computational package COMSOL 5.2. As the numerical model is highly nonlinear and thus unstable, the Poisson and Nernst-Planck equations are first solved with a surface charge of -0.01 µC/cm$^2$ and 1 mV to minimize the nonlinearity and increase stability. The solutions are then used as initial values in a second step. In this second step all three equations are solved with the aid of a sequential solver: first the PNP equations followed by the Stokes equation, and the values of the surface charge and potential are ramped to -2 µC/cm$^2$ and 1 mV using the auxiliary sweep in 1 V and 10 × µC/cm$^2$ steps. The ramping reduces the effect of the nonlinearity in the stability. The scaling factors of the solver are referenced to the initial values. The discretization of the laminar flow is set to P2+P1 ($2^{nd}$ order velocity, $1^{st}$ order pressure) and to quadratic for the concentration of species. In the second step the sequential solvers are set to 'automatic highly nonlinear'.

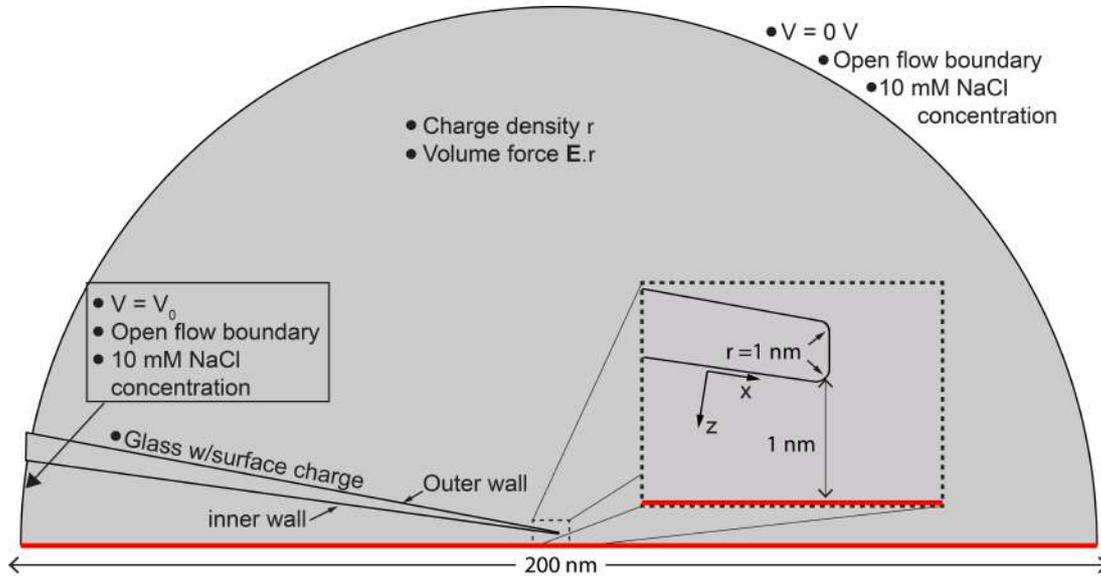

**Figure S6: COMSOL simulation.** Geometry, boundary conditions, and initial values used in the COMSOL simulation.

**Meshing**

The mesh is set to physics-controlled (fluid dynamics) and normal size everywhere but in the vicinity of the micro-capillary. A boundary mesh of 20 layers, with a first layer thickness of 0.01 nm and growth rate of 20%, was created along the outer surfaces of the whole micro-capillary. The mesh was further refined on the edges of the tip of the micro-capillary and at the corners of the glass with the boundary of the water. Here the maximum size element was set to 2 nm with a growth rate of 10%. This configuration ensured proper sampling around these sensitive areas. The mesh on the edges of the micro-capillary was set between these sensitive points to a maximum size of 40 nm, a minimum size of 1 nm, and a growth rate of 10%. The use of this mesh is vital to reach convergence of the solvers. In a 2.6 GHz Xeon processor, the calculation for all potential values lasted 55 minutes, consuming 3.88 GB and 5.01 GB of physical and virtual memory, respectively.

**Numerical results for the charge density and DC field**

Figure S7 shows the volumetric charge density in the vicinity of the inner and outer interfaces of the micro-capillary in the z direction normal to the interfaces. The plots are taken 40 μm from the tip.

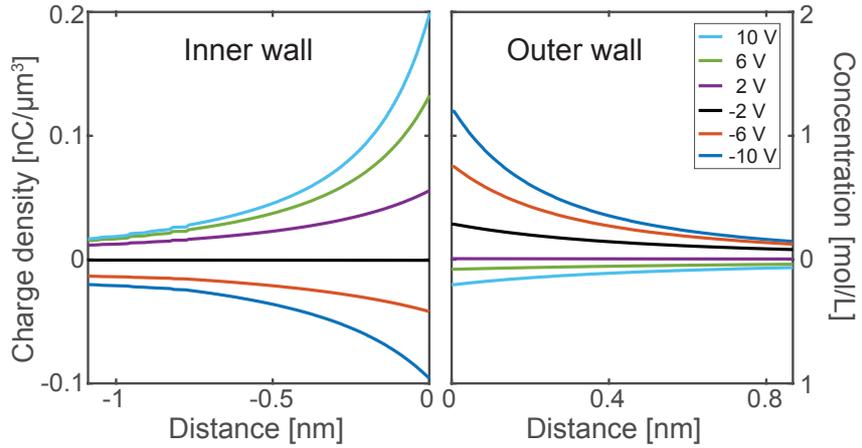

**Figure S7: Simulated volumetric charge density in the EDL.** Simulated charge density normal to the surface of the inner and outer walls of the micro-capillary when applying different potentials.

Assuming that the electrostatic field is uniform along the walls and only varying in the direction normal to the surface (z), we can use Gauss's law to find the electrostatic field normal to the surface ($E_{DC,z}$) as

$$E_{DC,z} = \int \frac{\rho(z)}{\varepsilon_0 \varepsilon_r} dz \tag{S16}$$

For the condition that no external bias is applied the computed electrostatic field decays exponentially as depicted in Fig. S8a. The velocity of the ions in the direction parallel to the interface as a function of distance away from it is given in Fig. S8b.

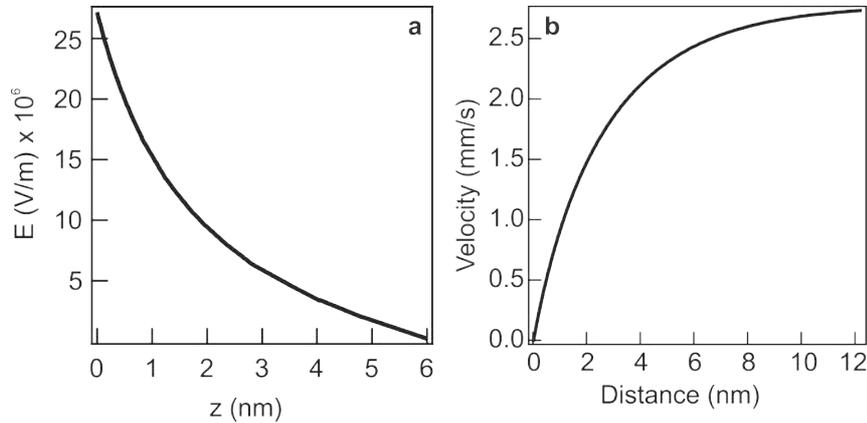

**Figure S8: Computed electrostatic field strength and ionic flow.** Electrostatic field strength at the inner wall as a function of distance for pH neutral conditions and 10 mM ionic strength. For the surface charge density a value of -2 μC/cm² was taken.

The value of the electrostatic field at the surface (z=0) of the inner and outer interface can be computed as well. To circumvent boundary issues in the simulation, we use the fact that the charge density (plotted in Fig. S7) decays exponentially with distance, and takes the form:

$$\rho(z,t) = A(t)e^{b(t)z} + C(t) \tag{S17}$$

The electrostatic field at the surface plane can then be written as:

$$E_{DC}(z=0,t) = \frac{A(t)}{b(t)} \tag{S18}$$

The values of the electrostatic field at $z = 0$ are plotted in Fig. S9 as a function of the applied bias (ΔV).

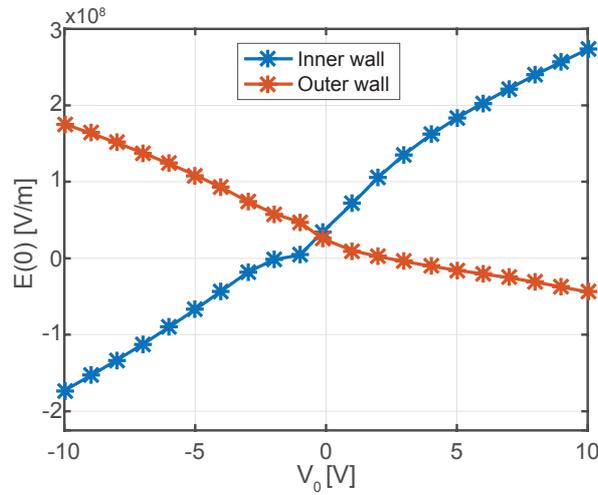

**Figure S9: Derived electrostatic field at the surface.** Electrostatic field at the surface of the inner and outer walls as a function of the applied potential in the electrodes derived from Eq. S17.

## S9. Modeling the SH data of Figure 3

Figure S10 shows the integrated simulated SH intensity in the XXX polarization direction from the micro-capillary when applying a 0.05 Hz voltage sine wave from -10 to 10 V. The intensity is calculated from Eq. S5 and adds the contributions from the inside and outside walls of the micro-capillary.

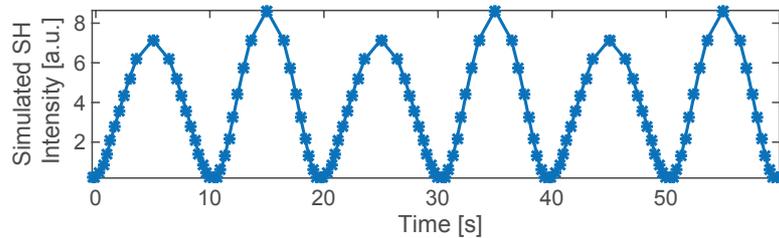

**Figure S10: Simulated integrated SH intensity.** Equation S5/1 is used to calculate the intensity from the inner and outer wall and these two contributions are added together.

Figure S11 shows a comparison of the measured intensity integrated across the whole image as a function of time (red) and the simulated SH intensity from Fig. S10 (blue). The external bias is shown in green. It can be seen that the spatially averaged intensity agrees with the computed result.

Figure S12 shows another comparison of the intensity obtained experimentally and in the simulations from the inside and outside walls of the micro-capillary. The solid lines depict the simulations results and the markers are the experimental points. The experimental points result from averaging 1x10 pixels (a 1.6 µm$^2$ area) of the images outside and inside the micro-capillary.

It can be seen that the simulation and experiment agree very well in Fig. S11, but less so in Fig. S12. The agreement in Fig. S11 is possible because both averaging the intensity across the whole image and the numerical simulations render a mean field steady state solution. Fig. S12 has more deviations because of the few points that are sampled and the large spatial and temporal fluctuations, in agreement with the analysis performed in S7.

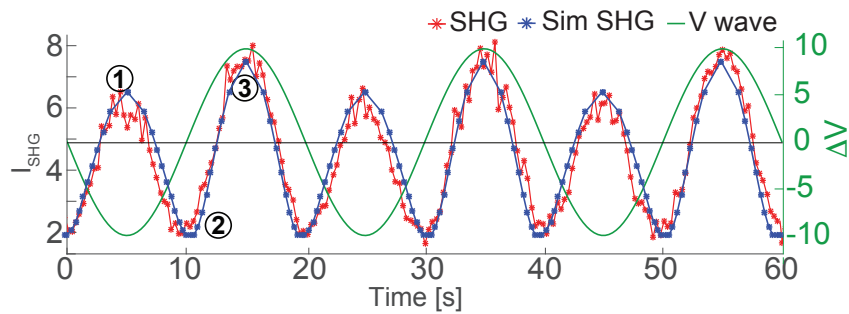

**Figure S11: Comparison of the simulated and measured SH intensity.** The measured intensity is integrated across the whole image as a function of time in red. The simulated SH intensity from Fig. S6 is shown in blue. The driving potential, which follows a 0.05 Hz sine wave from -10 to 10 V is shown in green.

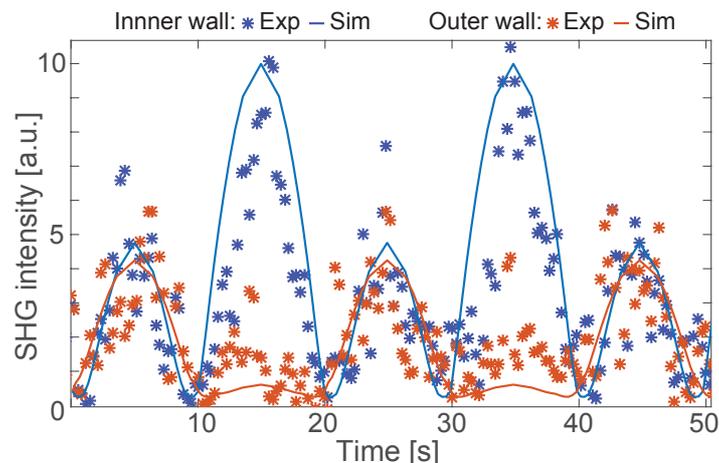

**Figure S12: Simulated and measured SH intensity.** The plots show the intensity from the inside (blue) and outside (red) walls of the micro-capillary. The solid lines depict the simulations results and the markers are the experimental points. The experimental points result from averaging 1x10 pixels on the images outside and inside the micro-capillary.